\documentclass[%
 reprint,
 amsmath,amssymb,
 aps
]{revtex4-2}

\usepackage{graphicx}
\usepackage{dcolumn}
\usepackage{bm}
\usepackage{xcolor}
\usepackage{float}
\begin{document}

\preprint{APS/123-QED}

\title{High speed and acceleration micrometric jets induced by GHz streaming: a numerical study with direct numerical simulations}

\author{Virginie Daru}
 \email{virginie.daru@ensam.eu}
 \affiliation{DynFluid Lab., Arts \& Métiers Science \& Technology, 151 boulevard de l'h\^opital, 75013, Paris, France}
\author{Michael Baudoin}%
 \email{michael.baudoin@univ-lille.fr}
\affiliation{Univ. Lille, CNRS, Centrale  Lille, Univ. Polytechnique Hauts-de-France, UMR 8520, IEMN, F59000 Lille, 
France }%
\affiliation{Institut Universitaire de France, 1 rue Descartes, 75005 Paris}

\date{\today}

\begin{abstract}

Gigahertz acoustic streaming microjets, with the capability of achieving fluid speeds up to meters per second, open new avenues for precision fluid and particle manipulation at microscales. However, theoretical and numerical investigations of acoustic streaming at these frequencies remain relatively scarce due to significant challenges including: (i) The inappropriateness of classical approaches, rooted in asymptotic development, for addressing high-speed streaming with flow velocities comparable to the acoustic velocity, and (ii) the numerical cost of direct numerical simulations generally considered as prohibitive. In this paper, we investigate high-frequency bulk acoustic streaming using high-order finite difference direct numerical simulations. First, we demonstrate that high-speed micrometric jets of several meters per second can only be obtained at high frequencies, due to diffraction limits. Second, we establish that the maximum jet streaming speed at a a given actuation power scales with the frequency to the power of 3/2 in the low attenuation limit and linearly with the frequency for strongly attenuated waves. Lastly, our analysis of transient regimes reveals a dramatic reduction in the time required to reach the maximum velocity as the frequency increases, following a power-law relationship of -5/2. This phenomenon results in remarkable accelerations within the Mega-g range at gigahertz frequencies. 

\end{abstract}

\maketitle


\section{\label{sec:int}Introduction}

Gigahertz (GHz) streaming is receiving increasing interest, thanks to the emergence of mature advanced technologies for generating acoustic waves at these frequencies and the myriad possibilities offered by high-speed micrometric jets for fluid and particle manipulation \cite{tac_duan_2023}.

Acoustic waves in the GHz range can be generated using various techniques, including Surface Acoustic Waves (SAWs), Lamb Waves (LW), or Bulk Waves (BW) resonators. High-frequency SAWs can be synthesized on the surface of piezoelectric crystals using miniaturized Interdigital Transducers (IDTs) \cite{arfm_yeo_2014}. In this case, acoustic energy remains localized at the substrate's surface, so that the use of thin films is not required. Shilton et al. demonstrated fluid actuation at GHz frequencies using SAWs in 2014 to achieve mixing in nanoliter droplets \cite{am_shilton_2014}. In the same year, Dentry et al. studied streaming jets generated by SAWs for frequencies up to 936 MHz \cite{pre_dentry_2014}. Additionally, Collins et al. reported highly localized acoustic streaming vortices in microchannels generated by focused IDTs at frequencies approaching the GHz range (636 MHz) \cite{ac_collins_2016}.

Generating high-frequency Lamb waves requires more advanced technologies due to the following factors: (i) The thickness of the piezoelectric layer used for wave synthesis must be on the order of or smaller than the wavelength to generate flexural or extensional modes, necessitating the use of thin films. (ii) To prevent the transmission of wave energy, the transducer must be isolated from its support. Consequently, high-frequency Lamb waves are typically synthesized using Lamb Wave Contour Mode Resonators (CMR), comprising a thin suspended piezoelectric film actuated with IDTs. High frequency CMRs have been introduced in the field of Acoustofluidics by Duan et al. \cite{acscs_duan_2018,prap_duan_2018}. But while they can operate in the GHz range \cite{ieee_zuo_2009}, their possibilities have not been explored so far in this regime for fluid actuation.

Finally, GHz bulk wave resonators are created by exciting longitudinal bulk modes in piezoelectric thin films sandwiched between two electrodes. To minimize losses in the supporting substrate, these resonators can either be suspended, as in the Thin-Film Bulk Acoustic Wave Resonator (FBAR) \cite{ieee_lakin_2003}, implemented on a Bragg reflector to form Solid Mounted Resonators (SMR) \cite{ieee_lakin_1995}, or deposited on the underside of a thicker substrate, which serves as a propagation medium and on top of which the microfluidics setup is attached \cite{ieee_ravi_2018}. It's worth noting that in the latter case, focused waves have been generated at GHz frequencies using Fresnel-type patterned electrodes, similar to the principle employed by Riaud et al. \cite{prap_riaud_2017} and Baudoin et al. \cite{sa_baudoin_2019} for synthesizing acoustical vortices.

With these BW GHz setups, high-speed jets reaching velocities of several meters per second, comparable to the acoustic velocity, have been reported \cite{pp_mettin_2015, npe_wu_2022}. These high-speed jets have been effectively utilized for various fluid manipulation tasks, including mixing \cite{apl_duan_2016,ieee_ravi_2018}, droplet dispensing \cite{loc_duan_2018}, and concentration as well as manipulation of micro- and nano-particles \cite{pps_duan_2018,nanos_duan_2019,sa_yang_2022,ieee_guo_2020,ieee_duan_2022} within microfluidic setups. It's important to note that, in contrast to classical miniaturized tweezers based on standing waves \cite{apl_tran_2012,pnas_ding_2012}, acoustic vortices \cite{prap_riaud_2017,sa_baudoin_2019,nc_baudoin_2020,arxiv_sahely_2022}, or pulsed methods \cite{collins2016acoustic,wang2021acoustic,chen2023numerical,kim2023acoustically}, which rely on the acoustic radiation force, manipulation in this case is accomplished using the Stokes drag induced by the jet \cite{tac_duan_2023}. The unique capabilities offered by these GHz acoustic manipulation techniques have paved the way for a new set of applications in biology, including controlled loading and release of vesicles \cite{ac_duan_2019}, modulation of neurite outgrowth \cite{lc_duan_2021}, and the manipulation of carbon dots to enhance bioimaging and biosensing \cite{t_zhang_2022}.

However, from a theoretical perspective, modeling high-speed acoustic streaming at GHz frequencies remains a challenging task because the streaming speed can be of the same order as the acoustic speed, rendering classical approaches based on asymptotic developments invalid \cite{arfm_baudoin_2020}. For instance, Duan et al. \cite{npe_wu_2022} employed a steady-state vs. harmonic decomposition approach, which, nevertheless, assumes that the steady-state components are much smaller than the harmonic components and is only suited for describing steady streaming. Steckel \& Bruus \cite{a_bruus_2020} modeled the entire problem of GHz streaming synthesis with Aln thin films, but only within the low-speed limit.

Since the early discussions by Stuart \cite{book_stuart} and Lighthill \cite{jsv_lighthill_1978}, there have been nevertheless various attempts to address the low speed limitation of streaming modeling. Gusev \& Rudenko \cite{spa_gusev_1979} investigated quasi-one dimensional streaming, considering hydrodynamic nonlinearities. Subsequently, Moudjed et al. \cite{pof_moudjed_2014} employed a time-scale discrimination approach, which segregates the field between fluctuating and time-averaged fields, reminiscent of the decomposition used in turbulence modeling. Using this approach, they derived scaling laws for the characteristic velocity of streaming jets as a function of applied power. Notably, Riaud et al. \cite{jfm_riaud_2017} adopted a similar approach to develop a unified model that encompasses nonlinear wave propagation, acoustic streaming, and radiation pressure, providing a simple expression for the Eckart Streaming source term. However, these authors simplified the acoustic streaming source term in the low-speed limit. More recently, Orosco and Friend \cite{pre_orosco_2022} proposed an original approach involving both time and spatial scale separation. Nevertheless, most of these works (i) are confined to relatively ideal configurations, (ii) lack comparison to direct numerical simulations or precise characterization of experimental velocity profiles, and (iii) have not been applied in the GHz range.

From a numerical perspective, the cost of Direct Numerical Simulation (DNS) of bulk acoustic streaming is generally considered prohibitive, and to the best of our knowledge, no such simulations have been reported in the literature so far.

In this paper, we employ a high-order finite difference code to perform direct numerical simulations of the Navier-Stokes compressible equations and compute transient and steady acoustic streaming. The simulations cover a wide range of parameters, including various transducer apertures and frequencies. Firstly, these simulations reveal that high-speed micrometric jets of several meters per second can only be achieved at high frequencies due to diffraction limitations. Secondly, the numerical simulations indicate that the scaling of jet speed with frequency transitions from a 3/2 power law to a linear power law as the attenuation length becomes comparable to the surrounding cavity, a trend rationalized with scale analysis. Finally, we leverage the unique opportunity presented by direct numerical simulations to investigate the transient development of acoustic streaming. This transient analysis demonstrates that at GHz frequencies, maximum velocities of up to tens of meters per second are  reached in just a few microseconds, resulting in tremendous accelerations on the order of $10^6 \; g$ (Mega-g). The physical insights provided by these numerical simulations establish a solid foundation for the development of appropriate theories for high-frequency streaming.

\section{\label{sec:int}Numerical method}

\begin{figure}[h]
	\centering
	\includegraphics[width=3in]{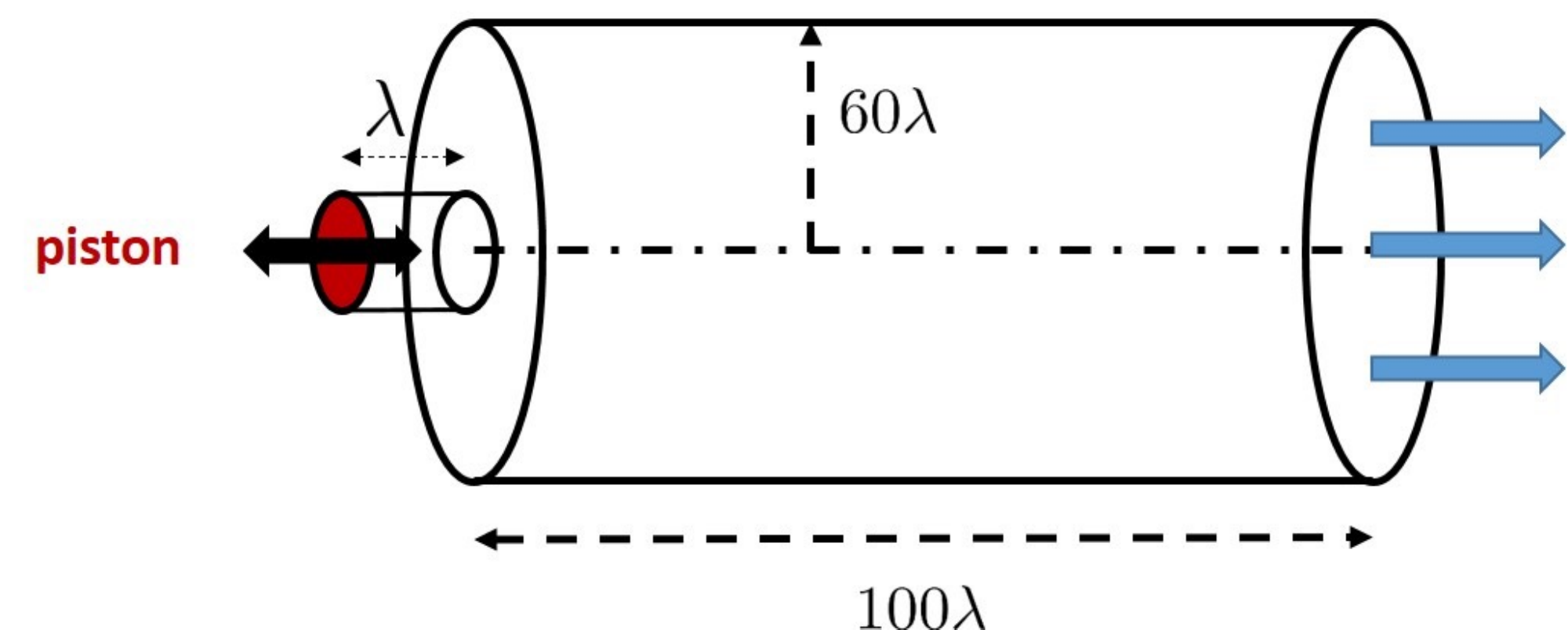}
	\caption{\label{fig1}Sketch of the geometry of the simulated problem. The piston has a radius equal to $R_p$.}
\end{figure}

The configuration studied here is represented on Fig. \ref{fig1}. A plane acoustic wave of limited aperture is generated by a cylindrical vibrating piston inside a liquid (water) initially at rest in a cylinder. The right end of the cylinder is open such that the flow is free to exit, while the lateral wall and the wall surrounding the piston are considered as solid walls. To perform simulations over a large range of frequencies, the dimensions of the simulation box (cylinder) are taken proportional to the wavelength and both the radius ($R_{cyl}=60 \lambda$) and length of the simulation box ($L_{cyl}=100 \lambda$) are chosen to be much larger than the acoustic source radius ($R_p$) to avoid interference of the radiated wave and jet with the walls. We consider piston radii $R_p$ between $0.5 \lambda$ and $8 \lambda$ and a wide frequency range varying from 37.5 Mhz to 3 Ghz. In all calculations, the injected power $P_{source}$ is kept constant, leading to a piston displacement amplitude proportional to $R_p/\lambda$, since the power is proportional to $(v_p R_p)^2$, $v_p$ being the piston velocity amplitude. The absolute value of $P_{source}$ was chosen to reach streaming velocity of the order of meters per second in the Ghz range in agreement with experiments \cite{pp_mettin_2015}.

The numerical approach consists of finite difference direct numerical solution of the isentropic compressible Navier-Stokes equations written in axisymmetric formulation:
\begin{equation}  \label{NS}
 \left\{  \begin{array}{l}
\displaystyle{ \frac{\partial \rho}{\partial t}+\nabla \cdot(\rho \mathbf{v})
}  =  0, \\[2mm]
\displaystyle{ \frac{\partial (\rho \mathbf{v})}{\partial t}+\nabla \cdot(\rho
\mathbf{v} \otimes \mathbf{v})+\nabla p
}  = \displaystyle \nabla \cdot \bar{ \bar{\tau} },\\
\end{array}
\right.
\end{equation}
with $\rho$ is the density, $\mathbf{v}=(v_r,v_z)$ the velocity, $p$ the pressure, $\bar{ \bar{\tau} }$
the viscous tensor. The bulk viscosity $\mu_B$ is taken into account, such that 
$\overline{\overline{\tau}}=2 \mu \overline{\overline{D}}
+(\mu_B-\frac{2}{3}\mu) \nabla \cdot {\bf v}  \overline{\overline{I}}
$, where $\mu$ is the shear viscosity, $\overline{\overline{D}} = \overline{\overline{\nabla}}^s \mathbf{v}$ the strain tensor, with $\overline{\overline{\nabla}}^s \mathbf{v}$ the symmetric part of the velocity gradient tensor.
Pressure and density are related by an isentropic equation of state of first order : $p-p_0=c_0^2 (\rho-\rho_0)$, with $c_0$ the speed of sound, the variables indexed 0 being the initial values. For water, the following values are used : $\rho_0=1000$ kg m$^{-3}$, $c_0=1500$ m s${-1}$, $\mu=10^{-3}$ kg m$^{-1}$ s$^{-1}$, $\mu_B=2.4\times 10^{-3}$ kg m$^{-1}$ s$^{-1}$. The system is operated at standard atmospheric pressure $p_0=101325$ Pa. A no slip Dirichlet boundary condition $\mathbf{v} = 0$ is applied on the lateral wall of the cylinder and around the piston. The vibration of the piston is simulated by applying an arbitrary Lagrangian-Eulerian (ALE)
method \cite{ALE} to the grid portion surrounding the piston, and the corresponding
grid portion is moving and deformable. The free flow in the outlet is simulated with a non reflecting boundary condition that is well approximated by a Neuman condition in our case (nearly plane wave). 

System (\ref{NS}) is solved using an explicit high order finite difference scheme (named OSMP) that was developed in \cite{OSMP} for compressible non viscous flows, and later extended for viscous flows (e.g. \cite{TAC2009}). It uses upwind formulae of order 7 for the convective terms, and centered formulae of order 2 for the viscous terms. The scheme is based on a coupled time and space (one-step) approach, resulting in the same formal order of accuracy in time and space for the convective part. The grid is Cartesian, uniform across the radius of the piston, stretched in the radial direction above, and uniform in the axial direction. The acoustic wavelength is discretized using 20 grid points per wavelength, that was proven to provide a sufficient accuracy for a good representation of the acoustic wave \cite{OSMPAIAA}. The scheme being explicit, the time step $\delta t$ is limited by the stability condition: $\frac{\delta t}{\delta x} c_0=\text{CFL} \le 1$, $\delta x$ being the size of the smallest cell.
This gives $\frac{T}{\delta t}=\frac{1}{\text{CFL}}\frac{\lambda}{\delta x} $, where $T=1/f$ is the time period. In our calculations we fixed $\text{CFL}=0.5$, such that 40 time iterations per time period were necessary. This results in a large number of iterations (several hundreds of thousands time steps) to reach steady streaming. Indeed, while the acoustic wave is very rapidly established in the domain, the streaming flow takes a much longer time to reach a steady state. Since direct numerical simulations are performed, the acoustic and streaming contributions are not naturally separated. To differentiate them and analyse the results, all variables $\phi$ are hence subsequently decomposed into average and fluctuating values $\phi=\overline{\phi}+\phi'$, such that the average value (corresponding to the streaming terms) is calculated from a simple arithmetic average over an acoustic time period of the variables $\overline{\phi} = \frac{1}{T}\int_0^T \phi(t) dt$ and the fluctuations (corresponding to the acoustic contribution) are simply obtained by subtracting the average field from the total field $\phi' = \phi - \overline{\phi}$. Note that since the field is averaged over a single period, this definition is perfectly compatible with the study of unsteady streaming.

\section{Diffraction induced limitations for the synthesis of microjets}

\subsection{Problem statement}
A natural idea to synthesize micrometric jets with high speed would be to concentrate the emitted power on a source of reduced (micrometric) dimension. In this section, we study how the streaming evolves as the radial dimension of the source is decreased.

\subsection{Results of the numerical simulations and analysis}

\begin{figure*}[htbp]
	\centering{
	\includegraphics[width=0.75\textwidth]{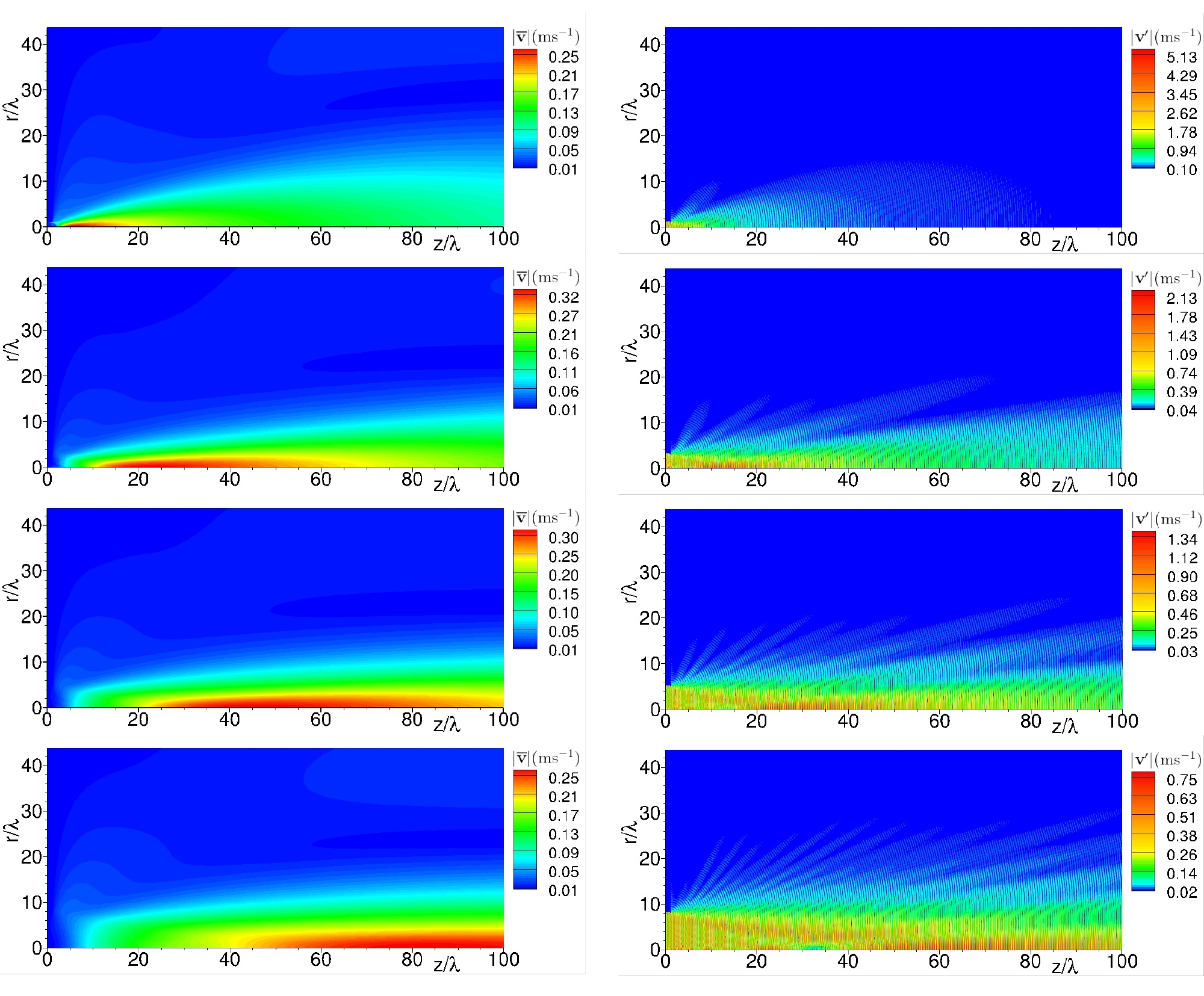}}
\caption{\label{fig2}Isolines of the modulus of the mean streaming velocity field $|\bar{\mathbf{v}}|$ (left) and acoustic velocity $| \mathbf{v}'|$ (right). From top to bottom: $R_p/\lambda=1,3,5,8$, $f=125$ Mhz.}
\end{figure*}

For this purpose, the frequency $f$ is fixed ($125$ Mhz) while the  piston radius $R_p$ is varied between $0.5 \lambda$ and $8 \lambda$. Figure \ref{fig2} represents the isocontours of the modulus of the streaming velocity field (left) and the acoustic velocity field (right), for four values of $R_p/\lambda$ (1,3,5,8). One can observe that the streaming velocity (Fig. \ref{fig2} left) is maximum in the vicinity of the axis in all cases. In addition: (i) The highest velocities are localized in the vicinity of the acoustic source in the case of $R_p/\lambda=1$, and then are shifted away from the source and spread along the axis when $R_p/\lambda$ increases. 
\begin{figure}[htbp]
	\centering{
	\includegraphics[width=3in]{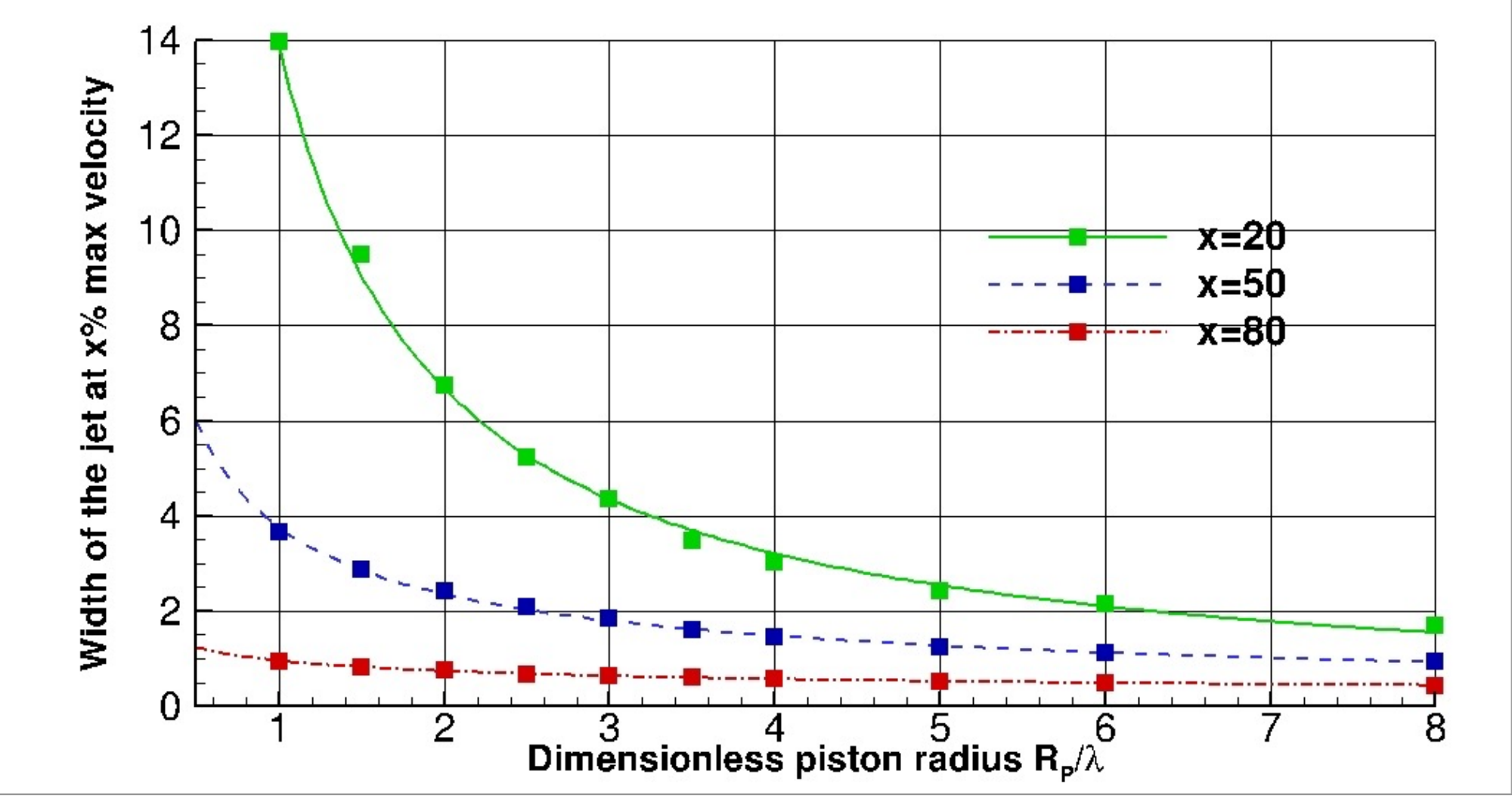}}
	\caption{Ratio of the radius of the jet defined as the isocontour corresponding to 20\% (green solid line), 50\% (blue dashed line), 80\%  (red dash-dotted line) of the maximum streaming velocity divided by $R_p/\lambda$, as a function of $R_p/\lambda$ at $f=125$ Mhz. The squares correspond to the numerical simulations and the lines to interpolation of the numerical points.}
 \label{fig3}
\end{figure}
(ii) The streaming jet is more and more concentrated around the axis  as $R_p/\lambda$ increases, as can be seen in Fig. \ref{fig3}. Finally, (iii) the maximum streaming velocity ($|\overline{\mathbf{v}}| \approx 0.34$ m/s) is obtained for $R_p/\lambda \sim 3$, and decreases for smaller values of $R_p/\lambda$ (Fig. \ref{fig4}).

All these effects result from an increase of the acoustic diffraction as $R_p/\lambda$ decreases (Fig. \ref{fig2}, right). Indeed diffraction (i) leads to the nearly cancellation, at a distance of a few wavelengths (for the smallest aperture) of the acoustic signal and thus of the source term responsible for the jet generation, (ii) spreads the acoustic signal and thus the streaming source terms laterally and (iii) prevents the possibility to concentrate the acoustic energy by reducing the transducer dimension to produce high speed streaming.
\begin{figure}[htbp]
	\centering
	\includegraphics[width=0.5\textwidth]{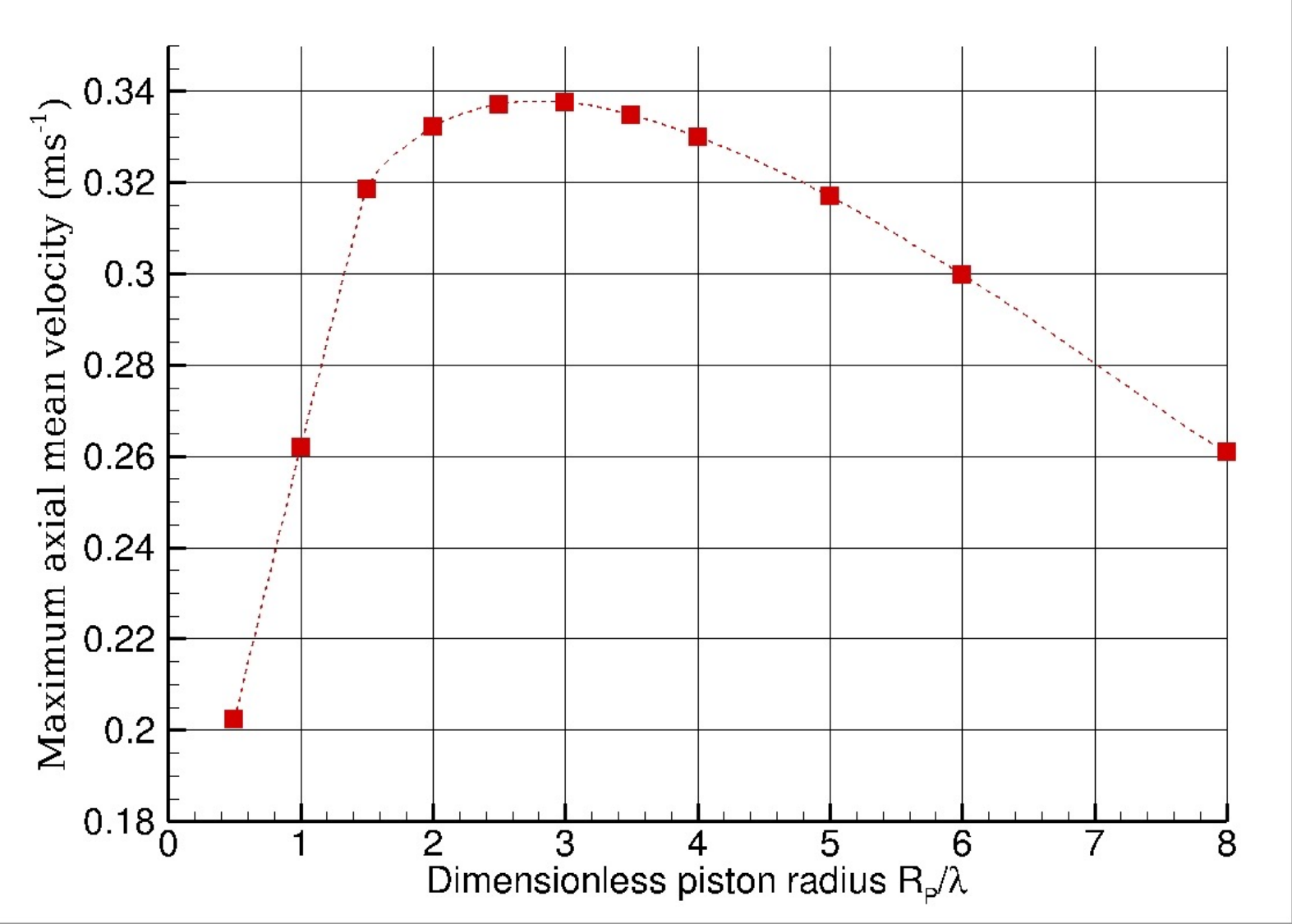}
	\caption{\label{fig4}Maximum axial mean velocity as a function of the dimensionless piston radius $R_p/\lambda$, $f=125$ Mhz. The red squares correspond the numerical simulations and the dashed dotted line to an interpolation of the numerical points.}
\end{figure}
\begin{figure}[htbp]
	\centering{
	\includegraphics[width=0.5\textwidth]{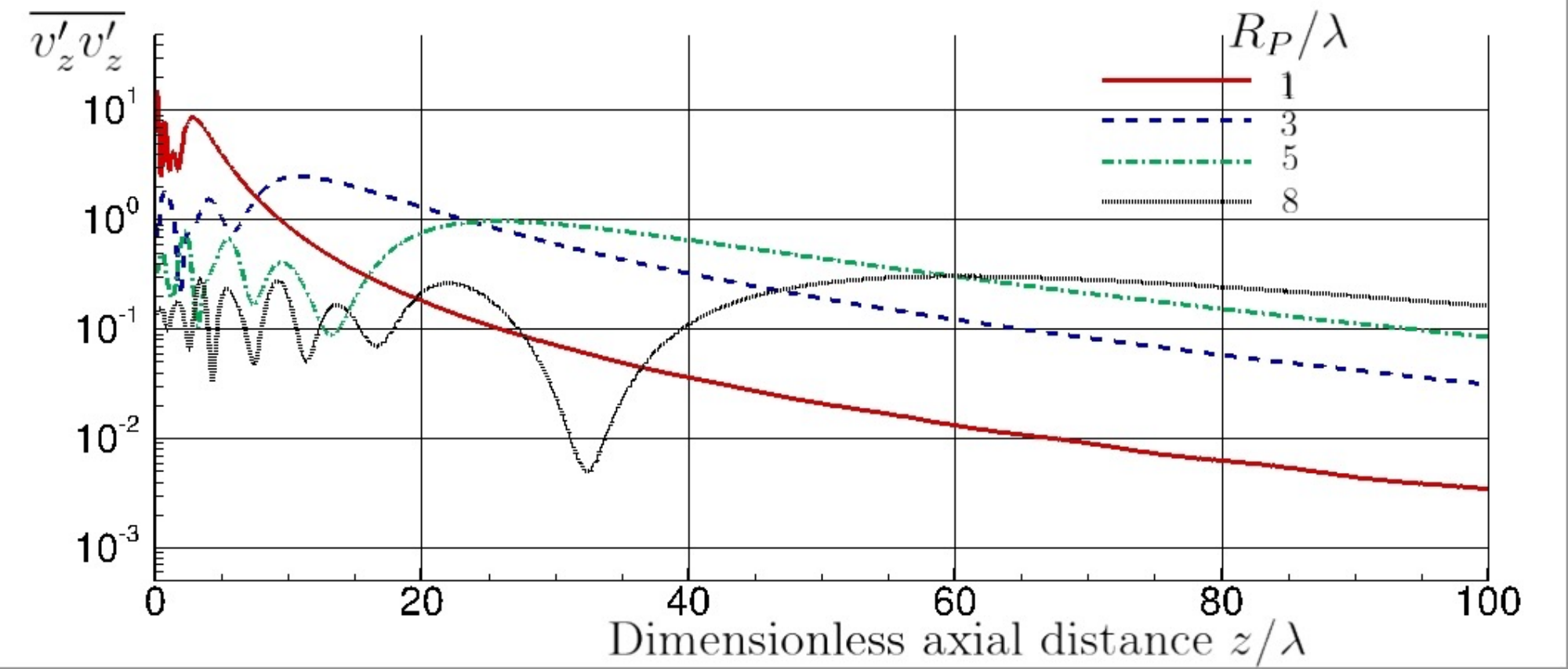}
	}
\caption{\label{fig5} Source term $\overline{v_z'v_z'}$ along the axis, in log scale, for $R_p/\lambda=1,3,5,8$, $f=125$ Mhz.}
\end{figure}
This cancellation of the streaming source terms at a few wavelength away from the source for the smallest aperture is clearly visible  on Fig. \ref{fig5}, which represents the dominant streaming source term \cite{pof_moudjed_2014} $\overline{v_z' v_z'}$ along the axis, in log coordinates. 

\begin{figure*}[htbp]
	\centering{
	\includegraphics[width=0.8\textwidth]{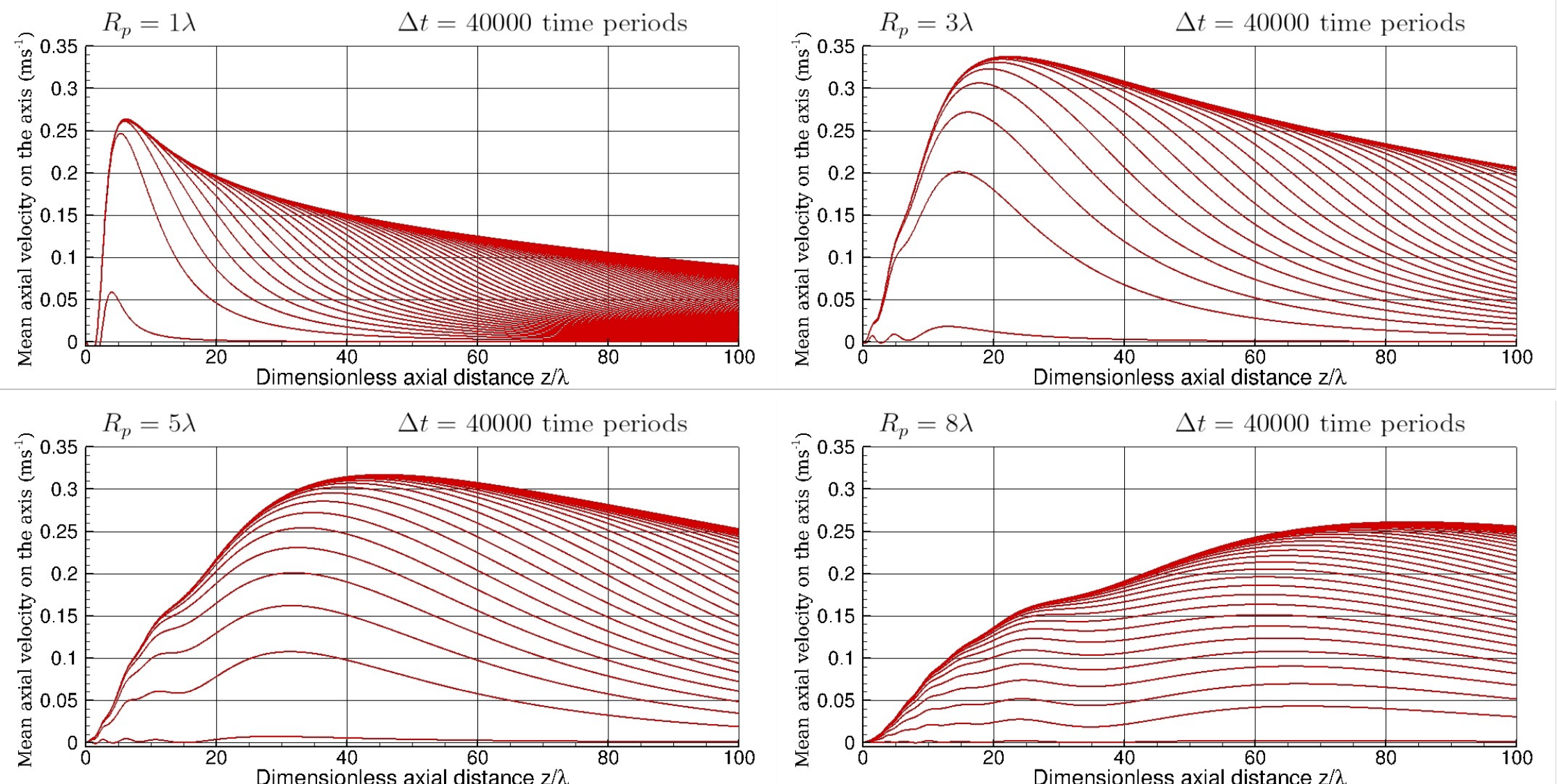}}
\caption{\label{fig6}Axial mean velocity along the axis every 40000 time periods, from top to bottom and left to right: $R_p/\lambda=1,3,5,8$, $f=125$ Mhz.}
\end{figure*}

To understand the genesis of acoustic streaming in these different configurations, it is also interesting to look at the temporal evolution of the streaming velocity along the axis, for different values of $R_p/\lambda$, which is made possible by our direct numerical simulations (Fig. \ref{fig6} and movies M1 and M2). In this figure, all curves are spaced by a constant time interval of 40000 time periods (corresponding to 0.32 ms). This figure shows that the jet is established in a different way for small and large values of $R_p/\lambda$. For small values ($R_p/\lambda=1$), the maximum velocity is rapidly reached near the acoustic source due to the localization of the source terms, and the velocity field is then transported by advection and/or diffusion. --- Note that in these simulations the jet Reynolds number $Re = \rho \, U D / \mu$, with $\rho$ the density of the fluid at rest, $U$ the jet characteristic velocity, $D$ its diameter and $\mu$ the dynamic viscosity is relatively constant and of the order of $Re \sim 30$, underlying that advection is expected to be dominant over viscous diffusion. --- For the largest value ($R_p/\lambda=8$), the velocity increases at approximately the same rate all along the axis, mainly due to the source term that is of the same order of magnitude everywhere. The advection effect should also be significant in this case, but the two phenomena are in this case entangled.

This section brings the conclusion that the only way to obtain a high speed localized jet is to increase the frequency to be able to concentrate laterally the streaming source without increasing diffraction effects.

\section{Frequency scaling of streaming jets}

\subsection{Problem statement}
Following this idea, we will now study the evolution of the acoustic streaming as a function of the actuation frequency. For this purpose the piston radius is now fixed to $R_p = 2 \lambda$ and the actuation frequency is varied from $37.5$ Mhz to $3$ Ghz.

\subsection{Results of numerical simulations and analysis}

\begin{figure}[htbp]
 \centering{
	\includegraphics[width=0.5\textwidth]{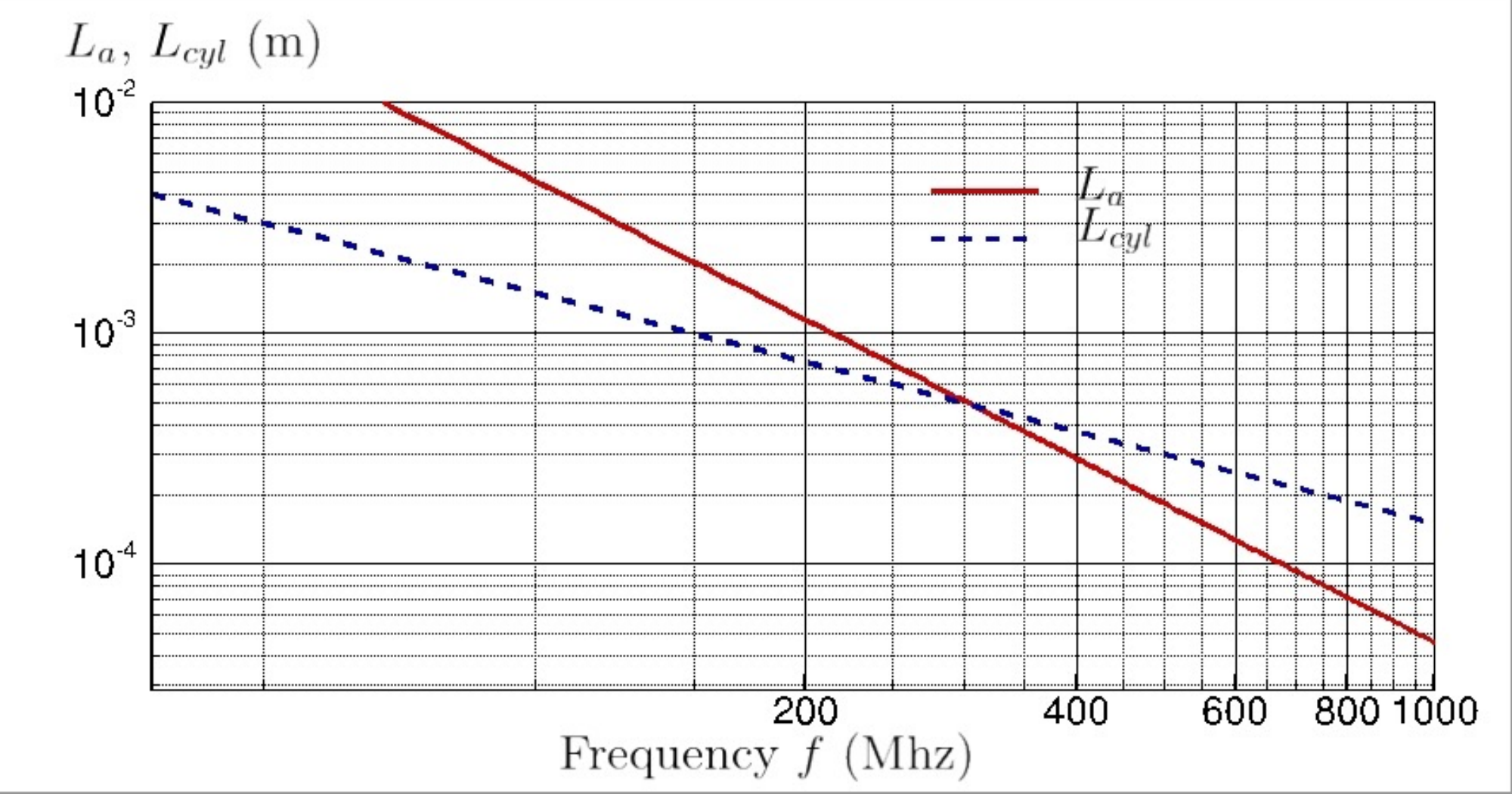}}
\caption{\label{fig7} Attenuation length $L_a$ (red continuous line) and cylinder length $L_{cyl}$ (blue dotted line) as a function of the frequency, in log scale.}
\end{figure}

Before delving unto the simulations, it is interesting to compare the acoustic wave attenuation length $L_a = \frac{2 \rho_0 c_0^3}{4 \pi^2 f^2 (\frac{4}{3}\mu+\mu_B)}$ with the simulation box length $L_{cyl}$ (Fig. \ref{fig7}). This figure shows, that below 300 MHz, the attenuation length is larger than the box length $L_{cyl}$, while above the attenuation length becomes smaller than the box length.

\begin{figure*}[htbp]
	\centering{
	\includegraphics[width=0.75\textwidth]{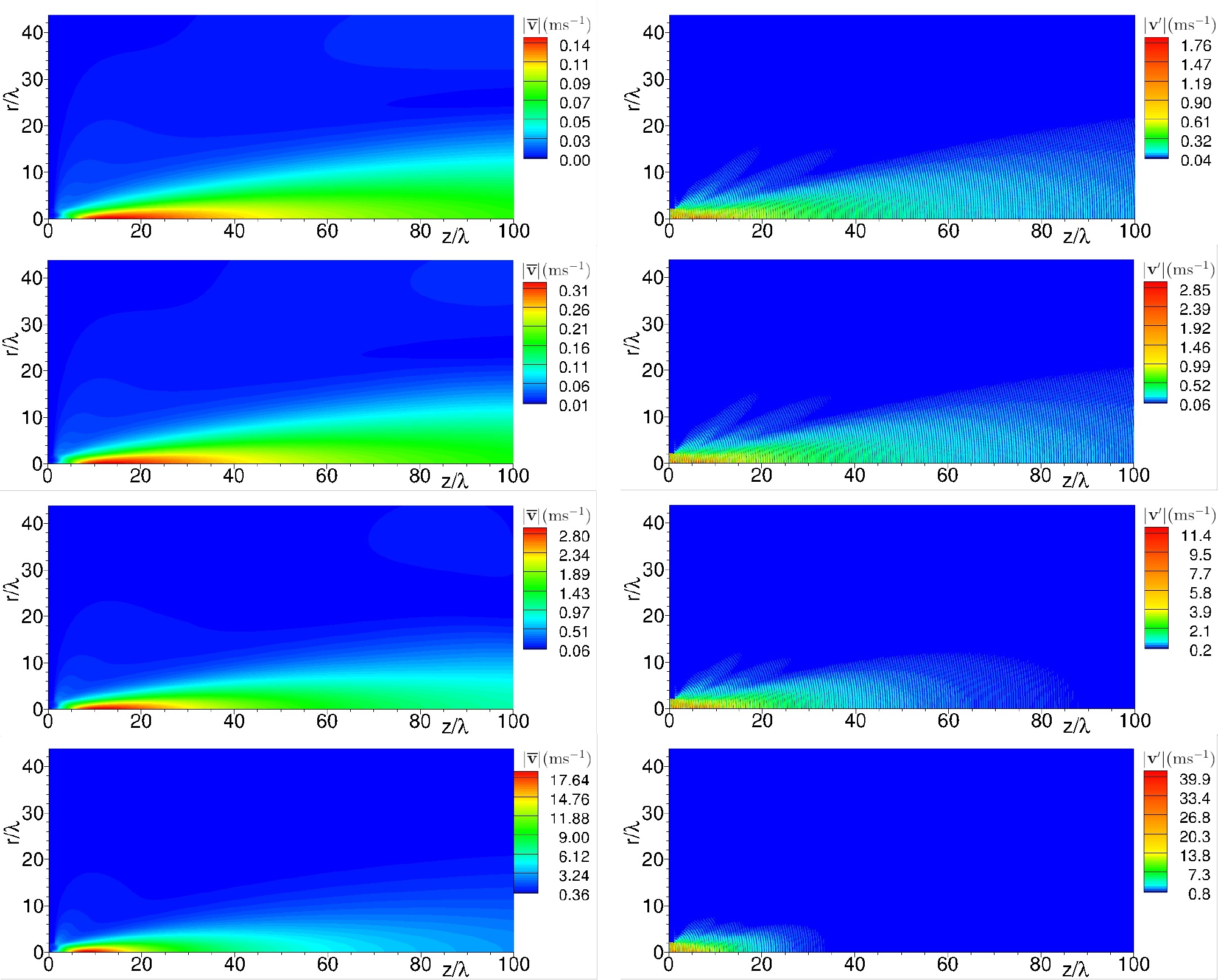}}
\caption{\label{fig8} Isolines of modulus of mean velocity field (left) and acoustic velocity (right), from top to bottom and left to right: $f=75,125,500,2000$ Mhz, $R_p/\lambda=2$.}
\end{figure*}

\begin{figure}[htbp]
 \centering{
	\includegraphics[width=0.5\textwidth]{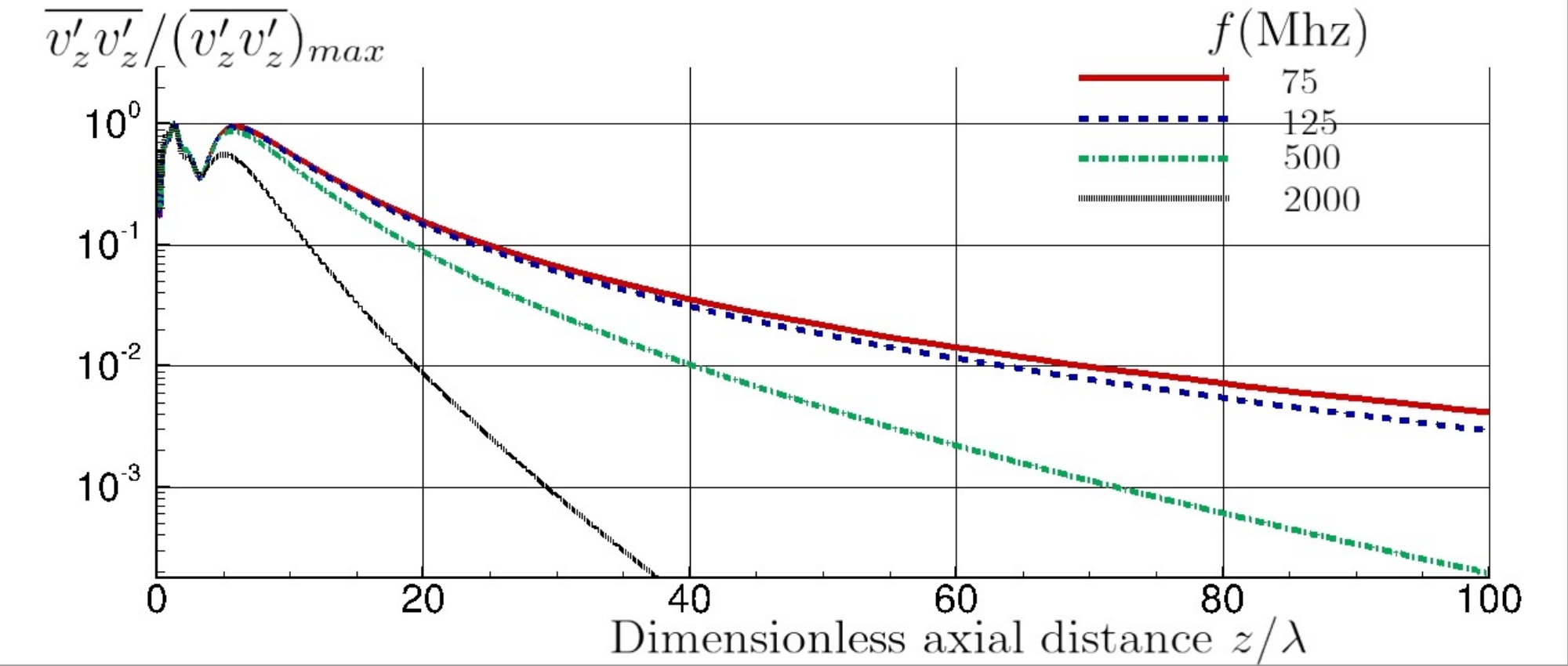}}
\caption{\label{fig9}Source term $\overline{v_z'v_z'}$ along the axis, normalized with the maximum value, from top to bottom and left to right: $f=75, 125, 500, 2000$ Mhz, $R_p/\lambda=2$.}
\end{figure}

\begin{figure*}[htbp]
	\centering{
	\includegraphics[width=0.8\textwidth]{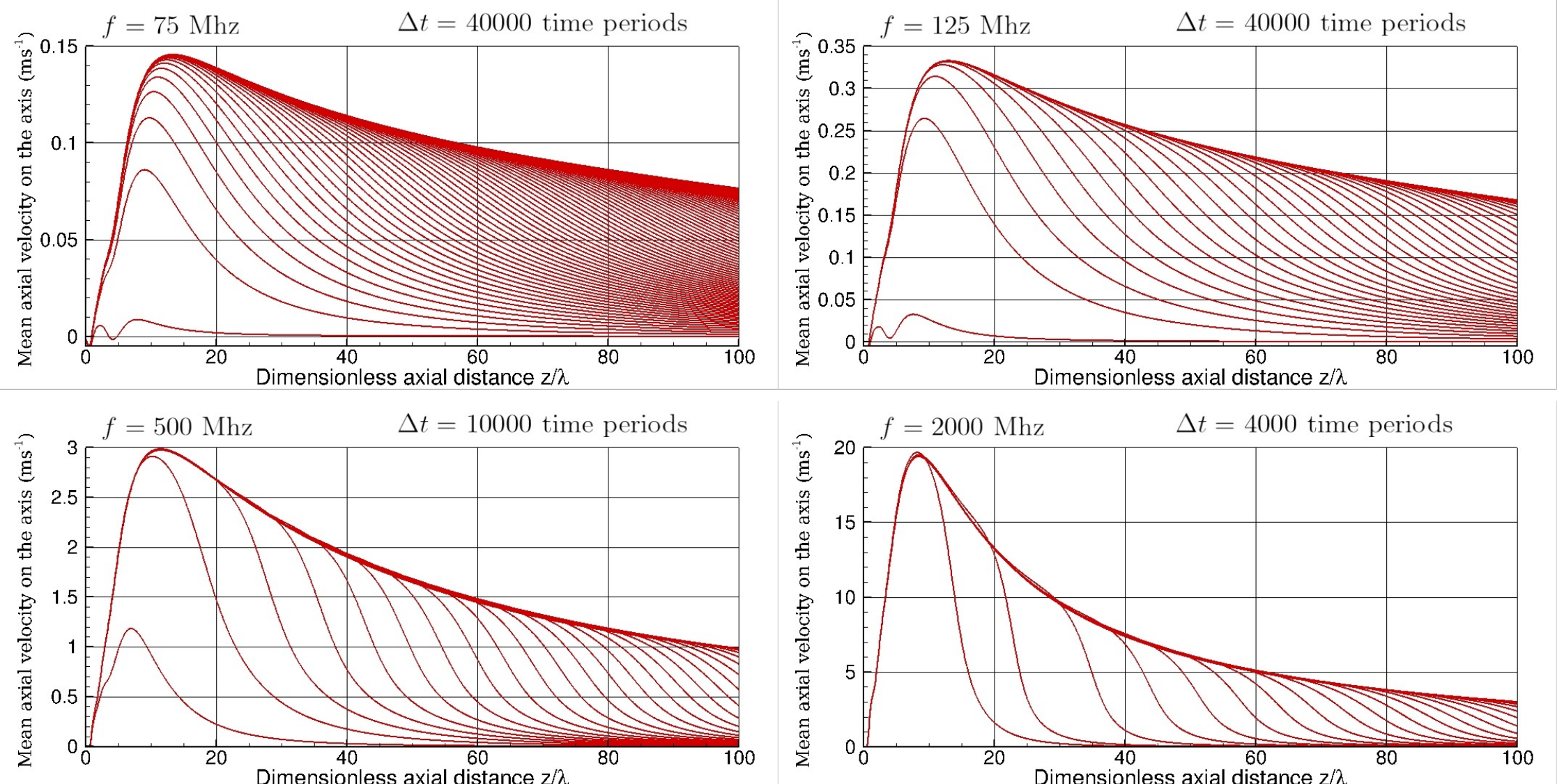}}
\caption{\label{fig10}Axial mean velocity along the axis every 40000 time periods, from top to bottom  and left to right: $f=75,125,500,2000$ Mhz, $R_p/\lambda=2$.}
\end{figure*}
 
The isocontours of the modulus of the streaming velocity field and the acoustic velocity field are represented on Fig. \ref{fig8} for four values of the frequency (75, 125, 500 and 2000 Mhz). This figure shows that the acoustic signal is rapidly attenuated for the highest frequencies leading to highly localized streaming source terms close to the source and jet structures modified accordingly with the highest speed in the close vicinity of the piston. The fact that in the case $L_{cyl} \ll L_a$, the fluid is forced in the entire domain while when $L_{cyl} \gg L_a$ the streaming source is localized in the vicinity of the source, can be seen from the plot of (i) the dominant streaming source term $\overline{v_z'v_z'}$ in log scale along the central axis (Fig. \ref{fig9}) and (ii) the temporal evolution of the average streaming velocity on the central axis (Fig. \ref{fig10}). This last figure shows that while the final velocity profile look relatively similar: (i) The ratio between 
the max and min velocity (the latter being obtained at the exit section of the cylinder) increases with the frequency due to the stronger localization of the forcing at high frequency. And (ii) the fluid is first accelerated close to the transducer over a distance $\sim L_a$ and then convected to larger z at high frequency while at lower frequencies, the fluid is accelerated simultaneously all along the axis.

\begin{figure*}[htbp]
	\centering{
 \includegraphics[width=0.8\textwidth]{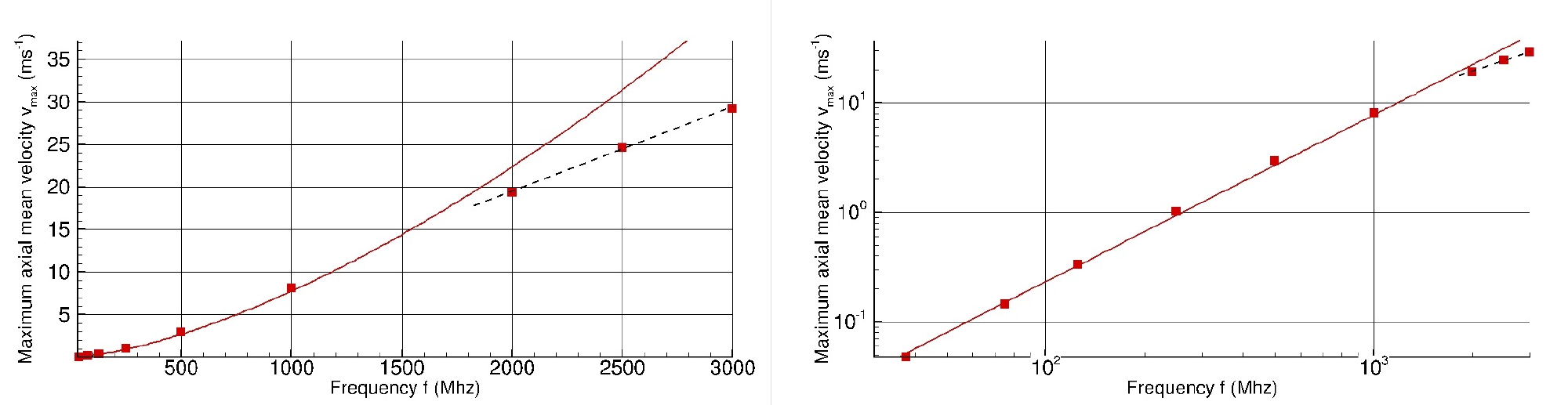}}
	\caption{\label{fig11}Maximum speed as a function of frequency (in linear scale on the left and log scale on the right). The red squares correspond to the value obtained with the direct numerical simulations. The red continuous line correspond to the best fit of the numerical points up to 1 GHz with a power law (leading to a coefficient of 1.52), while the black dotted line corresponds to a linear trend.}
\end{figure*}

\begin{figure*}[htbp]
	\centering{
 	\includegraphics[width=0.8\textwidth]{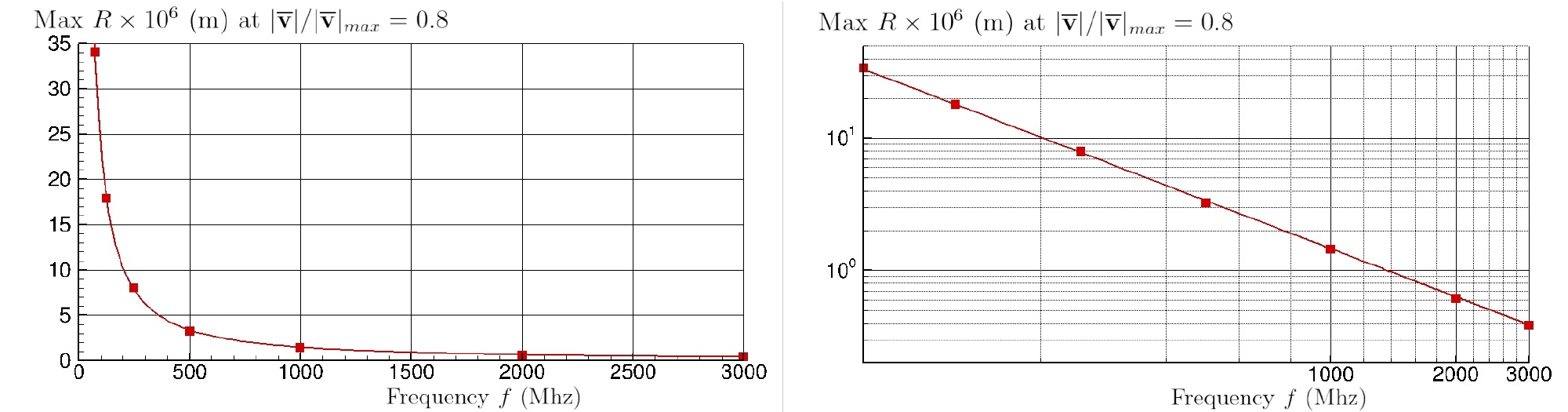}}
	\caption{\label{fig12}Maximal radius of the jet (in m) at a value of 80\% max velocity, as a function of $f$ (in log scale on the right). Red squares numerical simulations. Red continuous line, best fit with a power law (coefficient -1.2).}
\end{figure*}

The 3 next figures summarize the interest of high frequency sources to generate high speed microjets: (i) Fig. \ref{fig11} shows that the maximum jet speed scales as $f^{1.52}$ when $f \lesssim 1$ GHz, this dependence falling to linear when $f \gtrsim 1.5$ GHz. This tendency will be explained in the next section with dimensional analysis. This means that increasing the frequency enables to increase the jet speed. (ii) The radius of the jet (defined here as the radius of the isocontour corresponding $|v|/|v_{max}| = 0.8$ scales as $f^{-1.2}$ and thus almost as $\lambda$ for all frequency range (see Fig. \ref{fig12}). This can be  simply explained by the fact that here the ratio between the source radius and the wavelength $R_p / \lambda$ is taken as constant, and hence the diffraction is not limiting the scale reduction of the streaming jet, which is therefore proportional to the source radius. 

\subsection{Frequency scaling: a dimensional analysis}
\label{ss:fs}

Before producing a frequency scale analysis, it is interesting to study the evolution of the Reynolds number $Re$ introduced previously as a function of the frequency. In the present simulations, $\rho \sim 10^3$, and $\mu \sim 10^{-3}$ are fixed parameters and $U$ scales as $f^{1.5}$ below 1 GHz and as $f$ above (see previous section). Since the diameter of the jet is typically proportional to the wavelength $\lambda = c_0 / f$, it means that below GHz frequencies, the Reynolds number evolves slowly as the square root of the frequency $Re \propto \sqrt{f}$ and is not expected to evolve with frequency in the GHz range. Indeed, based on the simulations (Fig. \ref{fig8}), the diameter of the jet is of the order of $D = 20 \lambda \sim 250 \; \mu m $ at $125$ MHz and $D =5 \lambda \sim \; 7.5 \mu m$ at 2 GHz, while the velocity is of the order of $U\sim 0.15 \; m s^{-1}$ at 125 MHz and $U = 18 \; m s^{-1}$ at 2 GHz, leading to $Re \sim  40$ at 125 MHz and $Re \sim 135$ at 2 GHz. These Reynolds number values remain  far below the typical threshold value for the transition to turbulent jets of $Re \sim 2000$. This analysis indicates that even if high speed are reached at high frequency, the jet is not expected to become turbulent. They nevertheless indicate that the convective terms play an important role in this problem.

To obtain a scaling law for the evolution of the jet characteristic speed $U$ as a function of the frequency, we will start from the integral form of the kinetic energy conservation of the Navier-Stokes incompressible equations with a volumetric source term $\mathbf{f_S}$ corresponding to the streaming source:
\begin{eqnarray}
& & \iiint_v \rho_0 \left[ \frac{\partial e_c}{\partial t} + \mathbf{v}.\nabla e_c \right] dV \\
& & = \iiint_V - 2 \mu \overline{\overline{D}}:\overline{\overline{D}} \; dV  + \iint_{\partial V} \mathbf{\overline{\overline{\sigma_i}}}.\mathbf{v}.\mathbf{n} \, dS + \iiint_V \mathbf{f_S}.\mathbf{v} \, dV \nonumber
\end{eqnarray}
with $e_c = |\mathbf{v}|^2 / 2$ the kinetic energy density, 
$\overline{\overline{\sigma_i}} = -p \mathbf{I} + 2 \mu \overline{\overline{D}}$ the stress tensor in the incompressible regime
Here we look at the stationary phase so that the time derivative can be neglected. In the fluid acceleration region, the power produced by the streaming force is used to accelerate the fluid and produce kinetic energy. So if we balance the kinetic energy term and streaming source terms in the jet acceleration region,
$$
\iiint_v \rho_0  \mathbf{v}.\nabla e_c dV \sim \iiint_V \mathbf{f_S}.\mathbf{v} \, dV,
$$ we obtain:
$$
\frac{\rho_0 U^3}{L} \times \pi R_{jet}^2 L \sim U \int_{z=0}^{L_{cyl}} f_{s}(z) dz \times \pi R_{beam}^2
$$
with $U$ the characteristic velocity,  $L$ a characteristic length associated with the fluid acceleration driven by streaming, $L_{cyl}$ the simulation box size, $z$ the axial coordinate, and $R_{jet}$ and $R_{beam}$ the radius of the jet and beams respectively. From the simulations (Fig. \ref{fig8}), we can see that the radius of the jet is always comparable to the one of the acoustic beam $R_{jet} \sim R_{beam}$, which will be noted $R$ thereafter.  Then following Moudjed et al. \cite{pof_moudjed_2014}, the streaming force scales as $f_{s}(z) \sim \alpha I_{source}e^{-2 z / L_a} / c_0$ in the approximation of an attenuated quasi-plane wave, with $\alpha = 1/L_a$ the attenuation coefficient. If we introduce the acoustic power radiated by the source $P_{source} = \pi R_p^2 I_{source}$, 
$$
\rho_0 U^3 \sim \frac{P_{source} \alpha U}{c_0 \pi R_p^2} \int _{z=0}^{L_{cyl}} e^{-2 z / L_a} dz 
$$
i.e.
$$
U \sim \left[ \frac{P_{source} \alpha}{\rho_0 c_0 \pi R_p^2} \int_{z=0}^{L_{cyl}} e^{-2 z / L_a} dz \right]^{1/2}
$$
We can note that the scaling law obtained here is similar to the one obtained by Moudjed et al. \cite{pof_moudjed_2014} for near field streaming jets. But the kinetic energy integral formulation enables to clearly differentiate the two asymptotic regimes when the attenuation length is much larger than the simulation box length $L_a \gg L_{cyl}$ and the case when $L_a \ll L_{cyl}$. As discussed in the previous section and as shown on Fig.\ref{fig7} the transition between these two regimes occurs in the GHz range.  

When $L_a \gg L_{cyl}$, the fluid is accelerated by the streaming force all along the box, leading to $L \sim L_{cyl}$ and $\int_{z=0}^{L_{cyl}} e^{-2 z / L_a} dz \approx L_{cyl}$ since $e^{-2 z / L_a} \approx 1$. We hence obtain:
\begin{equation}
U \sim \sqrt{\frac{P_{source} \alpha  L_{cyl}}{\rho_0 c_0 \pi R_p^2}}
\label{eq:gg}
\end{equation}
Finally $P_{source}$, $\rho_0$ and $c_0$ are independent of the frequency, $\alpha$ scales as $f^2$ and both $L_{cyl} = 100 \lambda = 100 \times c_0/f$ and $R_p = 2 \lambda$ (Fig. \ref{fig8}) scales as $1/f$, so that:
$$
\boxed{
U \propto f^{3/2}
}
$$
When $L_a \ll L_{cyl}$, the fluid will be accelerated only on a distance comparable to the attenuation length so that $L \sim L_a$ and $\int_{z=0}^{L_{cyl}} e^{-2 z / L_a} dz \approx L_a/2$, leading to:
\begin{equation}
U \sim \left[ \frac{P_{source}}{2 \rho_0 c_0 \pi R_p^2} \right]^{1/2}
\label{eq:ll}
\end{equation}
since $\alpha L_a = 1$

This formula shows that when the attenuation length is smaller than the cylinder length in which the streaming jet is generated, the maximum fluid speed neither relies on the attenuation length (since all the power of the source is anyway dissipated) nor on the cylinder length $L_{cyl}$ since the acceleration is produced on a scale smaller than the cylinder length. Since $R_p^2 \propto 1/f^2$, we obtain:
$$
\boxed{
U \propto f
}
$$
This analysis shows that in this regime the jet speed increase as a function of the frequency results primarily from the reduction of the transducer size $R_p$. 

To summarize this section, the scale analysis predicts a transition from a power-law exponent of 3/2 to a linear evolution of the jet speed as a function of the frequency when the attenuation length becomes smaller than the cylinder length. This prediction aligns perfectly with the trends discussed in the previous section, as illustrated in Figure \ref{fig11}.

\section{From Deca- to Mega-g accelerations}

\begin{figure*}[htbp]
	\centering{
	\includegraphics[width=0.8\textwidth]{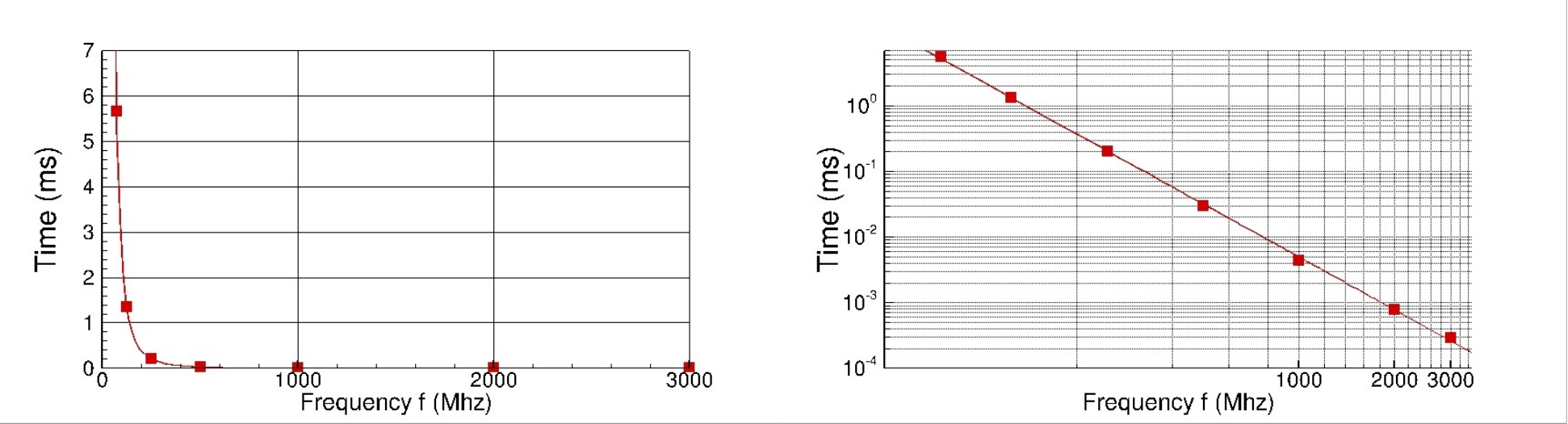}
}
	\caption{\label{fig13} Time necessary to reach the max velocity as a function of the frequency, in linear scale (left) and log scale (right). Red squares: numerical simulations. Red continuous line: best fit with a power law, leading to a coefficient -2.7.}
\end{figure*}

In this section, we explore an intriguing aspect of GHz streaming that, to the best of our knowledge, has not been previously documented in the literature but holds significant practical relevance for generating ultra-short and powerful jets. This aspect pertains to the remarkable reduction in the time $\tau_M$ required to reach the maximum jet speed as the frequency increases. As depicted in Fig. \ref{fig13}, our observations indicate that this time exhibits a scaling behavior inversely proportional to the frequency to the power of -2.7. This results in characteristic times of approximately 6 milliseconds at 75 MHz and merely 1 microsecond at 2 GHz. At the latter frequency, this acceleration corresponds to an order of magnitude of approximately 2 Mega-g. These trends in time reduction with increasing frequency can be  understood through scaling analyses in the unsteady case. Following subsection \ref{ss:fs}, we can balance the orders of magnitude of the unsteady term and source terms:
\begin{eqnarray}
\iiint_v \rho_0  \frac{\partial e_c}{\partial t} dV  \sim  \iiint_V \mathbf{f_S}.\mathbf{v} \, dV, \nonumber
\end{eqnarray}
leading to:
$$
\frac{\rho_0 U^2}{\tau_M} \pi R^2 L \sim \frac{P_{source} \alpha U}{c_0 \pi R_p^2} \int _{z=0}^{L_{cyl}} e^{-2 z / L_a} dz.
$$
This gives the following expression of the time $\tau_M$: 
$$
\tau_M = \frac{\rho_0 U \pi R_p^2}{\alpha P_{source}}
$$
Since $R_p^2$ scales as $1/f^2$, $\alpha$ as $f^2$ and $U$ scales as $f^{3/2}$ in the weakly attenuated regime and $f$ in the strongly attenuated regime, this scale analysis predicts that $\tau_M$ will scale respectively as $f^{-5/2}$ and $f^{-3}$ in these two regimes. No clear transitions between the two regimes is seen on Fig. \ref{fig13}, but the power $-2.7$ obtained from the best fit of the numerical points lies between these two limit values.

\section{On the limit of classical asymptotic developments.}

Finally, to conclude this article, it is interesting to compare the maximum streaming axial velocity $| \overline{\mathbf{v}}_{max} |$ to the piston source velocity $v_p$ and acoustic particular velocity $| \mathbf{v}' |$. 

\subsection{Streaming velocity vs piston velocity}

\begin{figure*}[htbp]
	\centering{
	\includegraphics[width=0.8\textwidth]{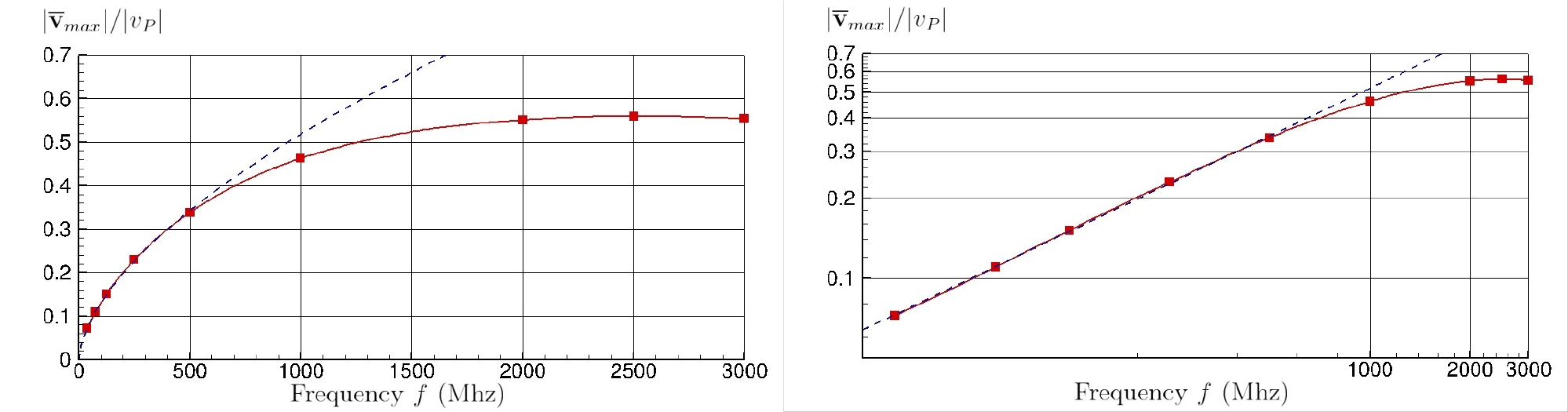}
}
	\caption{\label{fig14}Ratio of the streaming max velocity  $| \overline{\mathbf{v}}_{max} |$ divided by the the acoustic (piston) velocity $v_p$ as a function of the frequency (in linear scale at the left, and log scale at the right). The red squares corresponds to the numerical simulations. The red continuous line corresponds to an interpolation of the numerical points. The black dotted line corresponds to the best fit of the numerical points up to 500 MHz with a power law (leading to a coefficient of 0.6).}
\end{figure*}

Comparison to the piston velocity is provided on Fig. \ref{fig14}. This figure shows that $| \overline{\mathbf{v}}_{max} | / v_p$ scales as the square root of the frequency when $L_{cyl} \ll L_a$ and tends toward a constant when $L_{cyl} > L_a$. Indeed, $P_{source} = \overline{p_{source} v_{source}} \times \pi R_p^2$, where $p_{source}$ and $v_{source}$ correspond to the acoustic fluid velocity in the vicinity of the source. Due to the velocity continuity condition on the piston $v_{source} = v_p$, and assuming plane wave laws close the piston, $p_{source} = \rho_0 c_0 v_{source} = \rho_0 c_0 v_p$ leading to 
\begin{equation}
P_{source} = \rho_0 c_0 v_p^2 \pi R_p^2.
\label{eq:source}
\end{equation}
Now we can examine the two asymptotic limits. When $L_{cyl} \ll L_a$, we can combine eq. (\ref{eq:source}) with eq. (\ref{eq:gg}). We obtain
$$
\frac{U}{v_p} \sim \sqrt{\alpha L_{cyl}}
$$
a relation close to the one obtained by Orosco and Friend \cite{pre_orosco_2022} in the vicinity of the source, but here under an integral form. Now since $\alpha \propto f^2$ and $L_{cyl}=100 \lambda \propto 1/f$, we obtain:
$$
\boxed{
\frac{U}{v_p} \propto \sqrt{f}
}
$$
which is consistent with the tendency exhibited on Fig. \ref{fig13} (0.6 power law). When $L_{cyl} \gg L_a$, the velocity tends toward a constant limit which is obtained from the combination of eq. (\ref{eq:source}) and eq. (\ref{eq:ll}), i.e.
$$
\boxed{
\frac{U}{v_p}  \sim \sqrt{1/2} \approx 0.7
}
$$
Note that this asymptotic value was also obtained by Orosco and Friend \cite{pre_orosco_2022}. We see from Fig. \ref{fig14} that the streaming velocity is indeed saturating toward a value of the order of $0.6$. So the scale analysis gives even a quantitative approximation of the asymptotic limit.

\subsection{Streaming velocity vs acoustic particle velocity}

\begin{figure*}[htbp]
	\centering
	\includegraphics[width=\textwidth]{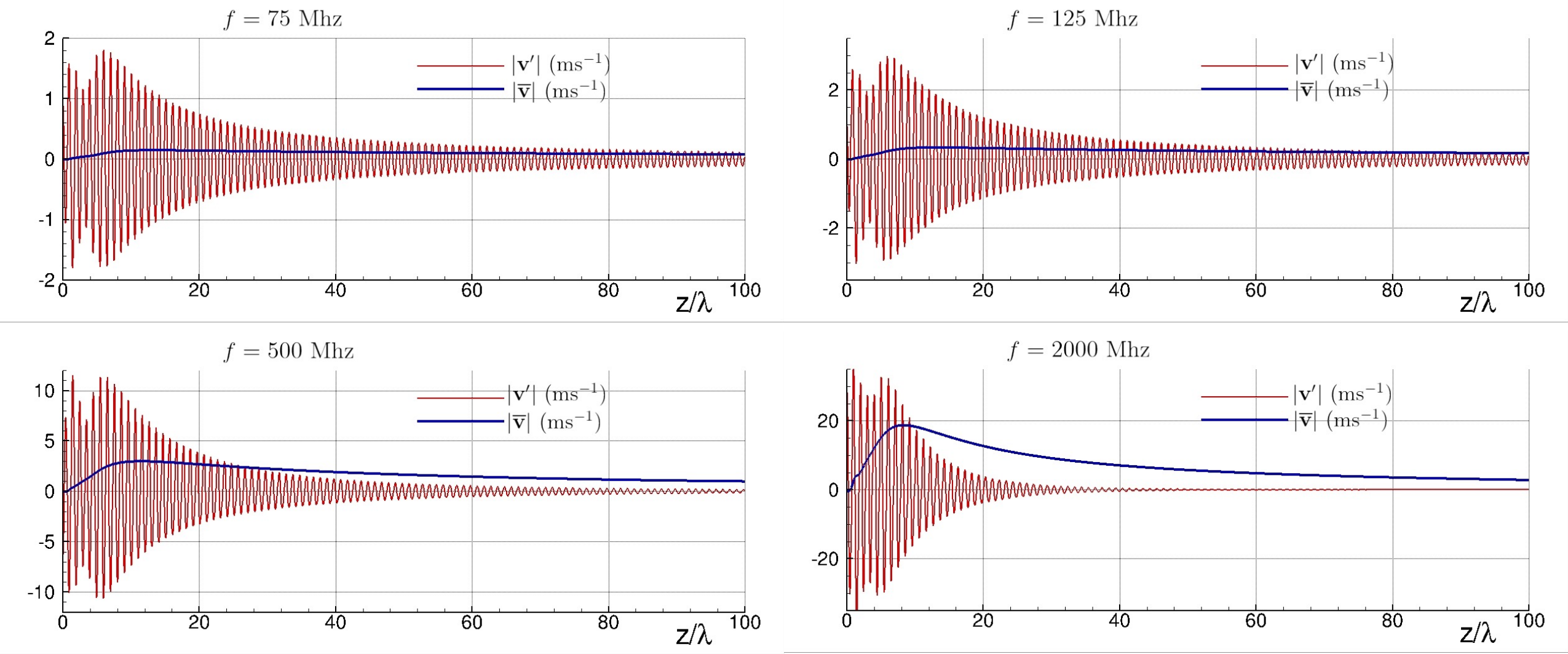}
	\caption{\label{fig15}Mean velocity (blue), and acoustic velocity (red) along the axis (in ms$^{-1}$), from left to right and top to bottom: $f=75,125,500,2000$ Mhz, $R_p/\lambda=2$.}
\end{figure*}

It is now interesting to compare the streaming velocity to the acoustic particular velocity. Fig. \ref{fig15} shows that at GHz frequencies the mean velocity can be of the same order as the acoustic particular velocity in the entire region giving birth to acoustic streaming. This figure underlines that classical asymptotic expansion stating that the streaming velocity is small compared to the acoustic velocity is no longer valid at these frequencies.

\section{Conclusion}

In this paper," we present the first direct numerical simulation of bulk acoustic streaming in a realistic configuration. By coupling these simulations with a scale analysis, we can elucidate the scaling law dependency of acoustic streaming on frequency and explain why micro-jets with speeds in meters per second can only be achieved at gigahertz (GHz) frequencies. This theoretical investigation also reveals a fascinating aspect of GHz acoustic streaming: achieving such high speeds could take just microseconds, resulting in remarkable acceleration in the Mega-g range. This aspect opens up new possibilities for generating ultra-short, high-speed microjets.

\begin{acknowledgments}
The authors acknowledge the support of the ERC Generator and Prematuration project programs funded by ISITE Université Lille Nord-Europe as well as funding from Institut Universitaire de France.\end{acknowledgments}

\end{document}